\documentclass{iau}
\usepackage{graphicx}
\usepackage{sidecap}
\usepackage{natbib}
\graphicspath{{./fig/}{./png/}}

\newcommand{\EQ}{\begin{equation}}
\newcommand{\EN}{\end{equation}}
\newcommand{\EQA}{\begin{eqnarray}}
\newcommand{\ENA}{\end{eqnarray}}

\newcommand{\Eq}[1]{Equation~(\ref{#1})}

\newcommand{\Fig}[1]{Figure~\ref{#1}}

\newcommand{\meanrho}{\overline{\rho}}

{}
{}
{}

{}
{}
{}
{}
{}
{}
{}
{}
{}
{}
{}
{}
{}
{}
{}
{}
{}
{}

{}
{}
{}

{}

{}
{}

\newcommand{\Rsun}{R_\odot}
\newcommand{\brac}[1]{\langle #1 \rangle}
\newcommand{\gggg}{\mbox{\boldmath $g$} {}}

\newcommand{\rr}{\mbox{\boldmath $r$} {}}

\newcommand{\UU}{\mbox{\boldmath $U$} {}}

\newcommand{\BB}{\mbox{\boldmath $B$} {}}

\newcommand{\JJ}{\mbox{\boldmath $J$} {}}

\newcommand{\AAA}{\mbox{\boldmath $A$} {}}

\newcommand{\ff}{\mbox{\boldmath $f$} {}}

\newcommand{\FF}{\mbox{\boldmath $F$} {}}

\newcommand{\nab}{\mbox{\boldmath $\nabla$} {}}

\newcommand{\SSSS}{\mbox{\boldmath ${\sf S}$} {}}

\newcommand{\DD}{{\rm D} {}}

\newcommand{\const}{{\rm const}  {}}

\def\Pm{\mbox{\rm Pr}_M}
\def\Rm{\mbox{\rm Re}_M}

\def\Rey{\mbox{\rm Re}}

\def\cs{c_{\rm s}}

\def\kf{k_{\rm f}}

\def\Brms{B_{\rm rms}}

\def\urms{u_{\rm rms}}

\def\Beq{B_{\rm eq}}

\def\half{{\textstyle{1\over2}}}

\def\onethird{{\textstyle{1\over3}}}

\newcommand{\Rc}{{R_{\rm C}}}
\newcommand{\yapj}[3]{ #1, {ApJ,} {#2}, #3}

\newcommand{\yapjl}[3]{ #1, {ApJL,} {#2}, #3}

\newcommand{\yan}[3]{ #1, {Astron.\ Nachr.,} {#2}, #3}

\newcommand{\yana}[3]{ #1, {A\&A,} {#2}, #3}

\newcommand{\ymn}[3]{ #1, {MNRAS,} {#2}, #3}

\newcommand{\ysph}[3]{ #1, {Solar Phys.,} {#2}, #3}
\newcommand{\yswsc}[3]{ #1, {JSWSC,} {#2}, #3}

\newcommand{\ypre}[3]{ #1, {Phys.\ Rev.\ E,} {#2}, #3}

\newcommand{\papj}[1]{ #1, {ApJ}, to be published}

\newcommand{\sapj}[1]{ #1, {ApJ}, submitted}

\title[Coronal influence on dynamos] 
{Coronal influence on dynamos}

\author[Warnecke \& Brandenburg]   
{J\"orn Warnecke$^{1,2}$\thanks{supported by the ERC AstroDyn
Research Project No.\ 227952.}
 \and Axel Brandenburg$^{1,2}$}
\affiliation{$^1$NORDITA, KTH Royal Institute of Technology and Stockholm University,
Roslagstullsbacken 23, SE-10691 Stockholm, Sweden, email: {\tt joern@nordita.org} \\
$^2$Department of Astronomy, Stockholm University,
SE-10691 Stockholm, Sweden\\
}
\pubyear{2014}
\volume{302}  
\pagerange{1--5}
\jname{Magnetic Fields--Throughout Stellar Evolution}
\editors{M. Jardine, P. Petit, \& H. Spruit, eds.}
\doi{??}

\begin{document}

\maketitle

\begin{abstract}
We report on turbulent dynamo simulations in a spherical wedge with
an outer coronal layer.
We apply a two-layer model where the lower layer represents the
convection zone and the upper layer the solar corona.
This setup is used to study the coronal influence on the dynamo action
beneath the surface.
Increasing the radial coronal extent gradually
to three times the solar radius and changing the magnetic Reynolds number,
we find that dynamo action benefits from the additional coronal extent in terms
of higher magnetic energy in the saturated stage.
The flux of magnetic helicity can play an important role in this context.
\keywords{MHD, Sun: magnetic fields, Sun: activity, Sun: rotation,
  turbulence, Sun: corona}
\end{abstract}
\firstsection 
\section{Introduction}
The solar magnetic field is produced by a dynamo operating beneath
the solar surface.
In the convection zone, the turbulent motions driven by convection
and shear from the differential rotation are able to amplify and
organize the magnetic field.
These fields manifest themselves at the solar surface in form of
sunspots, in which the field is so strong that the heat transported by
convection is suppressed, leading to dark spots on the solar disk.
One important feature of these sunspots is their latitudinally
dependent occurrence.
Averaging over longitude, one finds the typical behavior of equatorward
migration of the underlying mean magnetic field.
This behavior gives clear evidence for the existence of a dynamo
mechanism in the Sun.
In dynamo theory the $\alpha$-effect plays an important role, because
this effect describes the amplification of large-scale magnetic field
in the absent of shear.
In the Sun, it is believed that the $\alpha$-effect produces new
poloidal field from the toroidal field.
How strong its contribution for the production of toroidal field is,
is currently under debate \citep[see e.g.][]{KMCWB13}.

Numerical simulations of turbulent dynamos have shown that the
$\alpha$-effect can be catastrophically quenched at high magnetic
Reynolds numbers \citep[see][for a detailed discussion]{BS05}.
One possible loophole to alleviate the quenching is to allow for
magnetic helicity fluxes \citep{BF00,SB06,BCC09}.
In this context, it is very important to choose a realistic boundary
condition for the dynamo, which allows for magnetic helicity fluxes.
Preventing a transport of helicity out of the simulation domain may
influence the dynamo solution and the strength of the amplified
magnetic field.
Besides the magnetic helicity fluxes, commonly used boundary
conditions for the magnetic field such as vertical field or perfect
conductor restrict the dynamo and the magnetic field to certain
solutions.
This led to the development of the so-called ``two-layer model'',
where we combine the lower layer, in which the magnetic field is
generated by dynamo action, representing the solar convection zone
with a upper, force-free layer, representing the solar corona.

Our first application of the two-layer model led to the formation of
structures reminiscent of plasmoid- and CME-like ejections, driven by a forced
turbulent dynamo \citep{WB10,WBM11,WBM12}
and, subsequently, by a self-consistently
driven convective dynamo \citep{WKMB12} in the lower layer.
This indicates that the dynamo can be directly responsible for producing
coronal ejections and form structures in the solar corona.
But in the recent work of \cite{WKMB13}, the authors find that
differential rotation and the migration of the mean magnetic field can
be also influenced by the presence of a coronal layer.
In this paper, we investigate how the corona
influences the dynamo action.

\section{Model}

We use spherical polar coordinates, $(r,\theta,\phi)$.
The setup is the same as that of \cite{WBM11}, where we use a spherical wedge
with $0.7\Rsun\leq r\leq\Rc$, $\pi/3\leq\theta\leq2\pi/3$,
corresponding to $\pm30^\circ$ latitude, and $0<\phi<0.3$,
corresponding to a longitudinal extent of $17^\circ$.
$\Rsun$ is the radius of the Sun and $\Rc$ is the outer radius of the
coronal layer.
At $r=R$ the domain is divided at into two parts.
The lower layer mimics the convection zone,
where a magnetic field gets generated by turbulent dynamo action.
The upper layer is a nearly force-free part, which
mimics the solar corona.
We solve the following equations of compressible magnetohydrodynamics,
\begin{eqnarray}
{\partial\AAA\over\partial t}&=&\UU\times\BB+\eta\nab^2\AAA,\\
\label{eq:ind}
{\DD\UU\over\DD t}&=& -\nab h +\gggg + {1\over\rho} \left(\JJ\times\BB
  +\nab\cdot 2\nu\rho\SSSS\right)+\FF_{\rm for},\\
{\DD h\over\DD t}&=&-c_s^2\nab\cdot\UU,
\label{eq:ent}
\end{eqnarray}
where $\AAA$ is the magnetic vector potential and the magnetic field
is defined by $\BB=\nab \times \AAA$, which makes \Eq{eq:ind} obey
$\nab\cdot\BB=0$ at all times.
$\eta$ and $\nu$ are the magnetic diffusivity and the kinematic
viscosity, respectively.
$\DD/\DD t=\partial/\partial t+\UU\cdot\nab$ is the
advective derivative, $\gggg=GM\rr/r^3$ is the gravitational
acceleration, $G$ is Newton's gravitational constant, and $M$ is
the mass of the Sun.
We choose $GM/\Rsun\cs^2=3$.
$\JJ \times \BB$ is the Lorentz force and
$\JJ=\nab \times \BB/\mu_0$ is the current density,
where $\mu_0$ is the vacuum permeability.
The traceless rate-of-strain tensor is defined as ${\mathsf
  S}_{ij}=\half(U_{i;j}+U_{j;i})-\onethird\delta_{ij}\nab\cdot\UU$,
where the semi-colons denote covariant differentiation,
$h=\cs^2\ln\rho$ is the specific pseudo-enthalpy,
with $\cs=\const$ is the isothermal sound speed.
As in the work of \cite{WBM12}, the forcing function is only present
in the lower layer of the domain.
This means that the forcing function goes smoothly to zero in the
upper layer of the domain ($r\gg \Rsun$).
The function $\ff$ consists of random plane helical
transverse waves with relative helicity
$\sigma=(\ff \cdot \nab \times \ff)/k_{\rm f} \ff^2$ and wavenumbers
that lie in a band around an average forcing wavenumber of $\kf 
\Rsun\approx63$.
The forcing function also has a dependence on the helicity which is here chosen
to be $\sigma=-\cos\theta$ such that the kinetic helicity of the turbulence
is negative in the northern hemisphere and positive in the southern.
More detailed descriptions can be found in \cite{WBM11} and
\cite{Hau04}.
The magnetic field is expressed in units of the equipartition
value, $\Beq=\mu_0\urms\meanrho$, where
$\meanrho=\brac{\rho}_{r\le \Rsun,\theta,\phi}$,
$\urms=\brac{u_r^2+u_{\theta}^2+u_{\phi}^2}_{r\le
  \Rsun,\theta,\phi}^{1/2}$,
and $\brac{\cdot}_{r\le \Rsun,\theta,\phi}$ denotes an average over
$\theta$, $\phi$ and $r\le \Rsun$, i.e., over the whole dynamo in
region.
The fluid and the magnetic Reynolds numbers are defined as,
\begin{equation}
\Rey=\urms/\nu\kf,\quad \Rm=\urms/\eta\kf.
\end{equation}
Their ratio is expressed by the magnetic Prandtl number
$\Pm={\Rm/\Rey}$.

As an initial condition we use Gaussian noise as seed magnetic
field in the dynamo region.
Our domain is periodic in the azimuthal direction.
For the velocity field we use a stress-free boundary condition on all
other boundaries.
For the magnetic field we apply a perfect conductor conditions
in both $\theta$ boundaries and the lower radial boundary ($r=0.7\,\Rsun$).
On the outer radial boundary ($r=\Rc$), we employ vertical field conditions.
We use the 
{\sc Pencil Code}\footnote{\texttt{http://pencil-code.googlecode.com}}
with sixth-order centered finite differences in space and 
a third-order accurate Runge-Kutta scheme in time;
see \cite{Mitra09} for the extension of the {\sc Pencil Code} to
spherical coordinates.

\section{Dynamo action}

We perform 27 runs where we change $\Rc$ and $\Rm$, but keep $\Pm$ constant.
The letters for different sets indicate the coronal extents:
$\Rc/\Rsun=1$, 1.5, 2, 3, 1.2, 1.1, and 2.5 for Sets~A--F.
In the first four sets, we vary $\Rm$ from
$1.5$ to $220$, for the last three sets we use $\Rm=6$.

\begin{figure}[t!]
\begin{center}
\includegraphics[width=0.76\columnwidth]{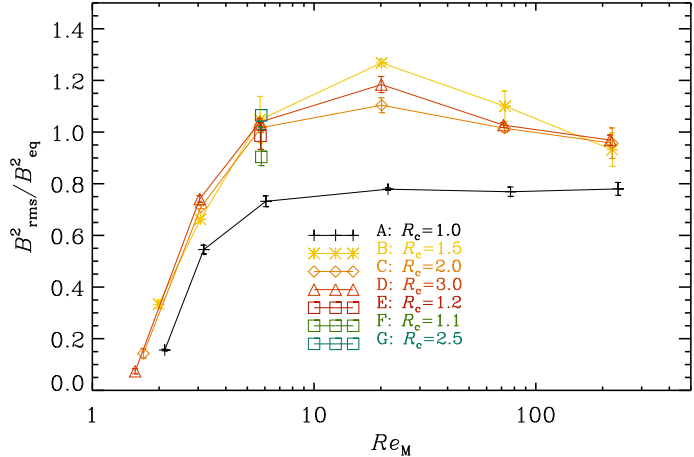}
\end{center}\caption[]{
Dependence of magnetic field energy normalized by the equipartition
value $\Brms^2/\Beq^2$ with coronal radial extent $\Rc$ and magnetic
Reynolds number $\Rm$. 
The solid black line indicates the dynamo region without corona.
}
\label{pcorona}
\end{figure}

For all runs the turbulent motion in the lower layer of the domain
drives dynamo action, which amplifies the magnetic field.
After exponential growth, the field saturates and shows cycles.
The field shows an equatorward migration of the all three magnetic
field components, as described in \cite{WBM11}.
This is caused by an $\alpha^2$ dynamo, where $\alpha$ changes sign
over the equator \citep{Mitra10}.
In \Fig{pcorona}, we show for all the 27 runs the normalized magnetic field
energy $\Brms^2/\Beq^2$ as function of magnetic Reynolds number $\Rm$.
The value for $\Brms^2/\Beq^2$ is obtained by averaging in space over the lower
layer of the domain $r\le\Rsun$ and in time over many hundred
turnover times in the saturated stage. 
The error bars in \Fig{pcorona} reflect the quality of the temporal averaging.
From \Fig{pcorona}, we can deduct two important results.
First, for runs with a corona the magnetic energy peaks at $\Rm\approx20$.
This seems to be not the case for runs without a corona.
On the other hand, the magnetic energy declines for larger $\Rm$,
as was also found by \cite{KKB10}, which could be related to a change
in the onset conditions for the different cases.
Second, the magnetic energies for all runs with a coronal extent are
larger by a factor of $\approx1.5$.
It seems that the actual radial size of the coronal extension is not
that important as long as there exists a coronal layer.
The run of Set~F has just a coronal extent of $\Rc=1.1\,\Rsun$, but
the magnetic energy is closer to runs with larger coronal extent than
to the one without corona.

Magnetic helicity fluxes might be a key to solving this riddle.
However, the outer radial boundary condition in the runs
without a corona also allow for magnetic helicity fluxes.
We recall that the simulation with a corona
generates large ejections of magnetic helicity \citep{WBM11,WBM12}.
Without possessing a coronal extent the dynamo might be not able to
produce ejection of magnetic helicity and therefore has a much lower
magnetic helicity flux through the boundary.
Studies on the nature of helicity fluxes in these runs are already on
the way (Warnecke et.\ al.\ 2013c, in preparation).
However, magnetic helicity fluxes might be important only much larger
magnetic Reynolds numbers \citep{DGB13}.

\section{Conclusions}

We have shown that a coronal layer on the top of a dynamo region can
support dynamo action.
This is visible through an increase in magnetic energy by adding a corona
at the top of the domain and leaving all other parameters the same.
However, it will be necessary to study magnetic helicity fluxes
through the surface of the lower layer for these cases to derive any
further conclusions.
The two-layer model has been used before to show the impact of a
dynamo on coronal properties and generating CME-like ejections
\citep{WB10,WBM11,WBM12}.
With this model is was also possible to generate spoke-like
differential rotation and equatorward migration in global convective
dynamo simulations \citep{WKMB13}, whereas models without a corona have
not been able to reproduce these features in the same parameter regime
\citep{KMCWB13}.
Besides dynamo models, this two-layer approach is successful in
combination with stratified turbulence in
producing a bipolar magnetic region \citep{WLBKR13} as a possible
mechanism of sunspot formation.

\end{document}